# A real-time hourly ozone prediction system using deep convolutional neural network

Ebrahim Eslami, Yunsoo Choi*, Yannic Lops, Alqamah Sayeed
Department of Earth and Atmospheric Sciences, University of Houston, TX 77004
*corresponding author, ychoi23@central.uh.edu

**Abstract:**
This study uses a deep learning approach to forecast ozone concentrations over Seoul, South Korea for 2017. We employ a deep convolutional neural network (CNN). We apply this method to predict the hourly ozone concentration on each day for the entire year using several predictors from the previous day, including the wind fields, temperature, relative humidity, pressure, and precipitation, along with in-situ ozone and $NO_2$ concentrations. We refer to a history of all observed parameters between 2014 and 2016 for training the predictive models. Model-measurement comparisons for the 25 monitoring sites for the year 2017 report average indexes of agreement (IOA) of 0.84-0.89 and a Pearson correlation coefficient of 0.74-0.81, indicating reasonable performance for the CNN forecasting model. Although the CNN model successfully captures daily trends as well as yearly high and low variations of the ozone concentrations, it notably underpredicts high ozone peaks during the summer. The forecasting results are generally more accurate for the stations located in the southern regions of the Han River, the result of more stable topographical and meteorological conditions. Furthermore, through two separate daytime and nighttime forecasts, we find that the monthly IOA of the CNN model is 0.05-0.30 higher during the daytime, resulting from the unavailability of some of the input parameters during the nighttime. While the CNN model can predict the next 24 hours of ozone concentrations within less than a minute, we identify several limitations of deep learning models for real-time air quality forecasting for further improvement.

## 1- Introduction:

Ambient ozone is an important phytotoxic pollutant. As one of the most harmful secondary air pollutants, surface ozone is mainly formed by the photochemical reactions between nitrogen oxides (NOx) and volatile organic compounds (VOCs) under certain meteorological circumstances [1]. With increased awareness of the health effects of ozone, having an accurate system for real-time ozone forecasting can be a significant benefit to public health by identifying adverse impacts of ozone [2]. In recent years, many research studies have focused on improving air quality models, specifically real-time ozone forecasting [3-6]. Real-time air pollution concentrations can be categorized by two models: deterministic models (e.g., chemical transport models) and statistical methods (e.g., machine learning techniques).

Air quality forecasting is commonly carried out via three-dimensional Eulerian chemical transport models, such as the United States Environmental Protection Agency (U.S. EPA)'s Community Multiscale Air Quality (CMAQ) model [7] to forecast extreme events and take necessary precautions to prevent extensive damage, especially in heavily populated areas. These models often report a significant model-measurement error that results from uncertainties in the treatment of physical processes and require higher run-time [6, 8-11]. Hence, statistical models, which are more computationally efficient, are also currently used for forecasting purposes. They include neural networks, regression methods, the fuzzy logic (FL) method, classification and regression trees (CART), and decision trees [6; 12-16].

One common class of statistical models is a neural network technique. The most popular of these models are multilayer perceptrons (MLP), recurrent neural networks (RNN), and radial basis function networks (RBFN) [17-18]. These techniques have been incorporated into multiple approaches for air quality forecasting [15, 19-21]. Certain concerns remain regarding the performance of such methods. For one, they require the determination of the optimum network structure (e.g., the number of hidden layers/units, input variables). In theory, non-linear physics can be approximated using a hidden layer with a large number of units [22], which can lead to "overfitting", necessitating the employment of an adaptive structure [23]. These methods also require a proper input dataset using suitable initial weights for accelerating learning convergence and avoiding stoppage at local error minima. Because of these limitations, the aforementioned "shallow" neural networks produce substandard results, necessitating the use of "deep" neural networks [24].

To utilize deep neural networks, Hinton et al. [25] developed a deep belief net (DBN) based on a fast, deep learning algorithm. In DBN, a layer-wise unsupervised learning algorithm is first used to pre-train the initial weights [25-26]. These weights are then fine-tuned by a global supervised learning approach with which more accurate architectures of hidden layers can be used for data-driven forecasting [18, 27]. Previous studies [18, 28-29] have indicated that for a set of nonlinear input variables, deep learning algorithms have reported higher forecasting precision than the conventional neural networks and the traditional autoregressive integrated moving average (ARIMA) time series model. The deep learning algorithm has been implemented in numerous applications in various fields such as finance, biology, and physics [30-32]. Kuremoto et al. [18] used a three-layer deep network to forecast a sample time-series. Hrasko et al. [33] implemented a hybrid neural network to predict time series obtained from three databases. Cai et al. [34] used a support vector machine (SVM) to apply a deep learning algorithm that facilitated regression analysis for stock price prediction. Li et al. [35] designed four deep learning models using feed-forward, energy-based, and recurrent architectures to predict the torsion angle of proteins. All of their models resulted in comparable accuracy.

For the application of this deep learning approach for weather and air pollution, Zhang et al. [29] applied a deep learning algorithm to predict short-term wind speed for up to two hours in one location. Li et al. [36] proposed a deep learning architecture for predicting $PM_{2.5}$ concentrations up to 12 hours using only previous concentrations. Despite achieving a relatively high prediction accuracy, the average mean absolute error of their predictions was around 9 $\mu g/m^3$, compared to the average concentration of 83 $\mu g/m^3$. Li et al. [37] utilized long short-term memory (LSTM) along with several machine learning techniques. Their results, however, showed a significant error for more than a four-hour prediction. Li et al. [38] applied a deep learning approach to improve the estimation of surface $PM_{2.5}$ using satellite data and surface observation.

In recent years, the convolutional neural network (CNN) [39], which was employed in this paper, has been acknowledged as the most successful and widely used deep learning approach [31]. The CNN is a biologically inspired, multistage architecture composed of convolutional, pooling, and fully connected layers that can be efficiently trained in a completely supervised manner. The key attribute of this model is to employ multiple processing units that can yield an effective representation of local data features. The deep architecture allows the stacking of multiple layers of these processing units that enable the characterization of data properties over several scales. Thus, the features extracted by the CNN are task dependent and "non-handcrafted."

Moreover, since the learning process of the CNN can be performed under the supervision of the output target, these features yield more discriminative power [40].

Given the computational efficiency of CNN in processing complex data, a CNN model can be potentially used as a real-time air quality forecasting system. However, real-time hourly ozone forecasting is challenging because of its non-linear chemistry and the highly varying and complex behavior of the atmosphere. Since ozone chemistry is significantly influenced by meteorology, which adds to the complexity of the problem, providing high prediction accuracy using conventional methods such as chemical transport modeling is challenging. This study introduces a deep learning technique for predicting hourly ozone concentrations for the entire year of 2017 over the city of Seoul, South Korea. Ozone is a secondary pollutant formed by reactions between primary pollutants such as $NO_2$. In Seoul, these pollutants are emitted by various sources and in various locations that are influenced by industry, automobiles, and biogenic sources [41-42]. This study uses a deep convolutional neural network to predict hourly ozone concentrations based on previous day observations of species and meteorological variables over multiple locations in Seoul.

**2- Methodology:**

Predictive conventional neural networks provide 24-hour forecasts based on a history of 24-hour observations (of the previous day). The backpropagation algorithm and gradient-based optimization technique are widely used for training such networks [22]. Deep networks (with more computation layers) with large initial weights usually lead to poor local minima. Those with small initial weights, however, produce shallow gradients in the later layers, which decrease the applicability of training networks with numerous hidden layers [26]. To resolve this issue, Bengio et al. [24] used a greedy layer-wise learning technique to train deep networks effectively. As the training strategy for this technique, the first layer learned the simpler concepts, and then the next layer learned more abstract features using the feature representation provided by the previous layer. Hence, the objective was to train the deep network layer-by-layer and use the backpropagation algorithm to fine-tune all of the network parameters [15, 24].

A common deep learning architecture for regression problems is the MLP, which uses the feedforward approach with backpropagation. Unfortunately, this approach results in inaccurate predictions of global error minima and consumes a great deal of computational time when dealing with a large size of input variables and non-linearity [24]. In addition, as hidden units can be very inefficient, especially in networks with multiple hidden layers [43], the learning process of earlier hidden layers could be significantly slower than that of latter layers [44]. Because of these issues, this study proposed using a deep CNN architecture. CNN is a common deep learning architecture that has long been applied in classification problems [39, 45-48]. Unlike other methods, CNN, capable of joint feature and classifier learning, can achieve greater classification accuracy on large-scale datasets [49-50]. Although numerous studies have applied CNN, very few [51-52] have used it for regression problems. To the best of our knowledge, this study is the first to adopt a deep regressive CNN approach for predicting hourly real-time air quality.

A schematic for the deep CNN used in this paper appears in Fig. 1. The figure shows that the CNN algorithm [53] has the input layer receives the time series of all input variables that are normalized to avoid a steep cost function. Each unit of a layer receives inputs from a set of units located in a small neighborhood in the previous layer. With local receptive fields, neurons can extract elementary features of inputs that are then combined with those of the higher layers. The outputs of such a set of neurons constitute a feature map (see Fig. 1). At each position, various types of units in different feature maps compute different types of features. A sequential implementation of this procedure for each feature map would be used to scan the input data with a single neuron with a local receptive field and to store the states of this neuron at corresponding locations in the feature map. The constrained units in a feature map perform the same operation on different instances in a time series, and several feature maps (with different weight vectors) can comprise one convolutional layer; thus, multiple features can be extracted at each instance [53]. Once a feature is detected, its exact "location" becomes less important as long as its approximate position relative to the other features is preserved [32].

The CNN captures changes in the temporal variation of the input data by sweeping through time series using a kernel of a given size. The various sections of the data are represented by feature maps. An additional layer performs a local averaging, called "pooling," and a subsampling reduces the resolution of the feature map and the sensitivity of the output to possible shifts and distortions. This step could potentially discard important information (e.g., sudden ozone peaks) [54] Hence, this study uses the convolution layer without pooling. The feature maps are connected to a fully-connected layer, which helps map each feature for multiple inputs to hourly ozone output (see Fig. 1). We use a deep convolutional neural network with five convolutional layers followed by a fully-connected layer before the output layer. Each convolutional layer is characterized by 32 filters and a kernel size of 2, while the fully-connected layer features 256 hidden units. We employed a rectified linear unit (ReLU) as the activation function in each layer applied to the normalized input data (since ReLU only passes values greater than zero). The algorithm was implemented in the Keras environment with the TensorFlow backend [55-56].

Compared to fully connected MLPs that have been extensively used as regression models, CNNs are attractive for several reasons. Firstly, MLPs are not explicitly designed to model variance within an ozone concentration that results from a complex interaction between several inputs. While MLPs of sufficient size could indeed capture invariance, they require large networks with a large training set. On the other hand, CNNs are more suitable for small-scale datasets than MLPs because they generally use a smaller number of parameters compared to fully-connected MLPs. Since input can be presented in any order without affecting the performance of the network, MLPs ignore input topology [53]. However, temporal variations of the ozone concentration have strong correlations, and modeling these local correlations with CNNs has been shown to be beneficial in other research areas [32].

To model the ozone concentration time-series, we used several predictors including hourly observed values of $O_3$ and $NO_x$ concentrations (as recorded by South Korea's National Institute of Environmental Research, or NIER), surface temperature, relative humidity, wind speed, and direction, dew-point temperature, surface pressure, and precipitation (as recorded by the Korean Meteorological Administration, or KMA). The location of air and meteorology stations can be found in Fig. S1 in the supplementary material.

This study used predictive deep learning techniques to forecast hourly surface ozone concentrations for the year 2017 and selected historical surface measurement data from 2014 to 2016 for training the model. Such a training period provides a broad history to fit a relationship between input variables and ozone concentration. For treating the missing observation data, SOFT-IMPUTE by Mazumder et al. [57] was applied to the raw measured data. SOFT-IMPUTE is a missing data treatment approach that iteratively replaces missing elements with those obtained from a soft-thresholded singular-value decomposition by taking all available data (spatially and temporally) into account. For a given station, we predicted each day's ozone concentrations in 2017 based on the observations from the previous day. At the end of the day of forecasting, we modified the weight matrix. We used 80% of the randomly selected data samples for training the model, and the remaining 20% for the validation process; the ratio was the result of a trial/error experience within the model configuration tuning. After each epoch, we monitored the performance of the model to make sure that the model stopped training at a minimum validation loss to avoid the possibility of overfitting.

### 3- Results and Discussion
### 3.1- General statistical analysis:

Table 1 compares the model-measurement statistics for all of the NIER stations in this study: the index of agreement (IOA), the correlation coefficient (r), the mean bias (MB), the mean absolute error (MAE), and the root mean square error (RMSE). The results of the deep learning technique demonstrate acceptable accuracy for the real-time prediction of the hourly ozone concentration for all stations. The yearly averaged IOA and r values of the CNN model were 0.87 and 0.79, respectively, while the values varied between 0.84 and 0.89 for the IOA and 0.74 and 0.81 for r. The yearly averaged MB, MAE, and RMSE were about 1 ppb, 9 ppb, and 12 ppb, respectively. This indicates that the CNN model, overall, underestimated the ozone concentration while missing very high-peak ozone episodes. Only three of the stations showed slightly positive MB (less than 0.3 ppb) while around half experienced negative MB of more than 1 ppb. The reason behind the relatively high MB is that the CNN model is unable to capture high-peak ozone concentrations specifically during the summertime.

The CNN model generally predicted the ozone concentrations of the stations located south of the Han River with higher IOA and r, and lower MB and RMSE (see the map plots of each statistical analysis in Fig. S2 in the supplementary material). The topology of the region south of the river is typically flat while that north of the region has a few elevated areas; thus, ozone formation in the southern region is more closely related to the variability of meteorological parameters than the northern region. The CNN model, therefore, is able to map a more accurate function to predict the ozone concentration in the southern region. In addition, the dominant wind direction is from the West and the Southwest (see the pollution-rose diagram of Fig. S3 in the supplementary material), both of which are influenced by the Yellow Sea. This dominant pattern results in a meteorology-dependent condition of the ozone concentrations in the southern part of the river and more variable ozone concentrations within the northern region.

Fig. 2 represents the daily prediction bias in different months of all 25 stations. We observed that the model bias varied more widely during the warm months (June-September) with more outliers. One explanation for this finding is that the wind pattern changed during these months (see Fig. S3) with relatively hot and humid days and occasional precipitation events. Most of the precipitation in Seoul occurs during the summer monsoon period between June and September, as a part of the East Asian monsoon season (see Fig. S4 in the supplementary material). Variable wind patterns along with scattered rain showers account for uncertainty in the CNN predictions during these months, leading to a larger bias range. Another explanation is that daytime ozone concentrations differed significantly from nighttime concentrations (see Fig. S6 in the supplementary material). This large difference caused a gap in the training process of the CNN model as it had less time to adjust the training process under such circumstances.

**Table 1.** Statistical analysis of the CNN ozone forecasting system for the entire year of 2017.

| Station ID | IOA* | r* | MB* (ppb) | MAE* (ppb) | RSME* (ppb) |
|---|---|---|---|---|---|
| 111121 | 0.85 | 0.76 | -1.41 | 9.90 | 13.30 |
| 111123 | 0.85 | 0.77 | -2.56 | 9.56 | 12.91 |
| 111131 | 0.86 | 0.77 | -0.45 | 7.56 | 10.40 |
| 111141 | 0.86 | 0.78 | -1.45 | 9.64 | 12.80 |
| 111142 | 0.87 | 0.77 | -0.38 | 7.86 | 10.78 |
| 111151 | 0.87 | 0.78 | -0.83 | 7.96 | 10.67 |
| 111152 | 0.88 | 0.80 | -0.57 | 7.44 | 10.07 |
| 111161 | 0.86 | 0.77 | -1.38 | 8.34 | 11.29 |
| 111171 | 0.87 | 0.79 | -0.53 | 9.17 | 12.23 |
| 111181 | 0.84 | 0.76 | -2.23 | 9.86 | 13.38 |
| 111191 | 0.85 | 0.74 | 0.19 | 8.76 | 11.43 |
| 111201 | 0.87 | 0.78 | -0.37 | 9.12 | 12.22 |
| 111212 | 0.88 | 0.81 | -1.31 | 9.71 | 13.34 |
| 111221 | 0.89 | 0.81 | -0.49 | 7.84 | 10.52 |
| 111231 | 0.87 | 0.81 | -0.78 | 8.80 | 12.28 |
| 111241 | 0.88 | 0.80 | -1.50 | 9.28 | 12.67 |
| 111251 | 0.89 | 0.81 | -0.31 | 8.65 | 11.69 |
| 111261 | 0.87 | 0.78 | 0.31 | 7.49 | 10.12 |
| 111262 | 0.87 | 0.80 | -1.38 | 8.87 | 11.88 |
| 111273 | 0.88 | 0.80 | -0.57 | 9.22 | 12.43 |
| 111274 | 0.88 | 0.80 | -1.10 | 8.89 | 12.05 |
| 111281 | 0.88 | 0.80 | -1.39 | 9.17 | 12.35 |
| 111291 | 0.85 | 0.76 | -2.18 | 10.93 | 14.46 |
| 111301 | 0.88 | 0.79 | 0.30 | 8.08 | 11.04 |
| 111311 | 0.87 | 0.79 | -1.12 | 10.48 | 13.99 |
| Average | 0.87 | 0.79 | -0.94 | 8.90 | 12.01 |

* IOA is the index of agreement, r is the Pearson correlation coefficient, MB is the mean bias, MAE is the mean absolute error, and RMSE is the root mean squared error.

### 3.2- CNN model prediction performance analysis:

The daily variations of the IOA and r of the CNN forecasting system are illustrated in Fig. 3. The figure shows that the variability of both parameters was higher during the cold months as well as in July. During the winter, most of the input parameters, the ozone concentrations, and the weights inside the CNN model that were trained to capture these extreme examples are at their lowest levels. During the cold months of the winter, the low (or near minimum) values in the weight matrix accounted for variable prediction performance. For July, however, the reason behind the variable prediction performances was the higher ozone concentrations during that month. Maximum concentrations in nearly one-third of the days in July 2017 exceeded 90 ppb level – the highest of all months in 2017. The outliers, in Fig. 3, affected the general prediction performance. One example representing an outlier in the CNN prediction appears in Fig. S5 in the supplementary material.

Fig. 4 shows the time-series comparison of daily mean and daily maximum ozone concentrations between the observations and the CNN predictions. The CNN successfully captured a trend in ozone for the entire year for all stations despite the noticeably different ozone trends (see Fig. 4(a)). For example, the ozone concentration at stations 111131(the third station from top left) and 111311 (the last station at the bottom right) followed different trends throughout the year 2017. Nevertheless, the CNN model predicted acceptable, similar trends for both stations (see Table 1).

The CNN model, however, significantly underpredicted the maximum daily ozone concentrations during high ozone months (see Fig. 4(a)). The model also "mispredicted" (predicting high maximum ozone concentrations even though observations revealed relatively low concentrations, and vice versa) concentrations on many days during the warm months because of the rapid change in weather conditions (see Fig. S7 in the supplementary material for more details). Since we trained the CNN model based on the information of the previous day, the model was unable to capture the aforementioned change. Nevertheless, for the training for the CNN model, we did not use several important meteorological parameters such as cloud fraction and solar radiation, which could represent the mentioned weather changes; continuous measurements of these parameters were unavailable for the city of Seoul for the time period of this study.

The results of the monthly IOA for all NIER stations is shown in Fig. 5. They show that the model was generally consistent in its predictions throughout the year. For each station, however, model performance varied from month-to-month. The reason for this variation was the availability of only one meteorological station (KMA station #108) in the city of Seoul for use as input, indicating that the meteorological inputs of the model were the same for all stations, resulting in variability in the prediction performance. In addition, we used only one ozone precursor ($NO_2$) as the emission representation for the input parameters, indicating that the CNN model had to rely on the quality of meteorological data from one meteorology station for the entire city.

### 3.3- CNN model diagnosis:

Although the CNN model can successfully predict hourly ozone concentrations with reasonable accuracies and within less than a minute of processing time, several issues deserve further investigation. One limitation of this study was the underprediction of the peak ozone. As illustrated in Fig. 4(a), the CNN model underpredicted all observed ozone concentrations over 80 ppbv. This large ozone bias could be attributed to the local emissions or the meteorological parameters that were not incorporated into the model. These biases could be mitigated by using big data (e.g., using a significantly larger period for training the model) combined with deep learning. By using big data, a more efficient learning environment (including more training examples, more input variables, larger input size, and so on) in which to efficiently train the deep learning algorithm would be available [58]. Another shortcoming of this approach is the lack of standard procedures for determining an optimal network architecture (e.g., the number of hidden layers/units, learning parameters), which is typically determined through trial and error and can greatly affect the performance of the model. As these hyper-parameters have internal dependencies, tuning them is prohibitively costly [45].

Another characteristic of CNN, like any statistical approach, is data sensitivity. Data sensitivity indicates that the quality of output directly depends on the input parameters. In light of the experience presented in this study, in which the ozone concentrations during the daytime and nighttime were separated, the performance of the CNN model differed. The monthly IOA of NIER stations calculated during the daytime and the nighttime are shown in Fig. 6. The differences between the IOA values for the daytime and the nighttime ranged from 0.05 to more than 0.3. The reason behind this notable difference is related to the difference in the values of input parameters during the day and night. All inputs (except for the relative humidity, $NO_2$, and the wind field) were at their daily minimum levels. Because of the normalization process, these values were usually close to zero in the input data. Thus, the model relied only on the relative humidity, $NO_2$ and the wind field for its predictions, which led to a significant reduction in the model performance.

For instance, the model had a significant underprediction during the nighttime when the wind blew from the South, while it generally overpredicted the ozone for the same wind directions during the daytime (see Fig. S8 in the supplementary material for more details).

Another drawback of this study is that we trained the CNN model using a limited number of variables from the previous day. Therefore, unlike the physical models, the CNN model was unable to take a physical phenomenon into account unless a proper indicator was used within the inputs. For example, Fig. 7 shows the mean bias of the CNN prediction in a categorical comparison using the relative humidity and the surface temperature. In a hot (high temperature) and humid (high relative humidity) condition, the model showed a significant overestimation (red areas in Fig. 7). This condition could have reduced the rate of the ozone formation and expressed by the presence of cloud fraction, a parameter that is lacking in the current CNN model. By contrast, a hot and relatively dry (lower than average relative humidity) condition can contribute to ozone formation during the daytime with the presence of its precursors (from local sources). In this condition, the model noticeably underestimated the ozone (blue areas in Fig. 7) owing to the lack of information about the local sources.

Deep learning extracts meaningful features of the raw input data through a greedy layer-wise learning process [24]. Hence, this algorithm develops a layered architecture to represent data and illustrates the influence of each feature [59]. A deep learning model, however, can only be trained with historical data for its feature extraction process. Thus, the sensitivity of the input parameters of the CNN model to the output might have been imbalanced because of the significant difference in the historical examples, one of which appears in the supplementary material (Fig. S9), indicating that the trend in ozone concentration changed with the wind direction from 2014 to 2017. Such anomalies in the trend can result in imbalanced prediction sensitivity at different levels of input parameters. Fig. 8 shows the annual mean of the CNN prediction at various category levels of four input parameters: wind speed, temperature, $NO_2$, and relative humidity. Because of the aforementioned limitation, the model prediction with respect to a change in the category level is less sensitive than the observation on a same day of the week. If we consider the monthly mean bias, we can find a similar phenomenon (see Figs. S10-12 in the supplementary material).

**4- Conclusions**

In this paper, we applied and trained a deep learning algorithm, deep convolutional neural network (CNN), and compared it with observation data. Deep CNNs can be trained to approximate smooth, highly non-linear functions [30], rendering them appropriate for forecasting non-linear processes related to air quality [15]. In addition, feature extraction using deep learning algorithms is more efficient than using conventional statistical methods, particularly when multiple hidden layers are structured [24].

We discussed the application of deep CNNs for the real-time prediction of ozone concentrations in Seoul, South Korea and trained the model to predict hourly ozone concentrations for the next day using the observations of $NO_x$, ozone and meteorological variables from the previous day. We evaluated the model for the entire year of 2017. This work has shown that the deep learning approach can predict hourly concentrations with sufficient accuracy (IOA=0.84-0.89, r=0.74-0.81) by modeling the relationship between local meteorological and species concentrations in an urban environment. The CNN model, which showed consistent prediction results across the city of Seoul, reasonably predicted daily and monthly trends of ozone concentrations throughout the year. However, the model generally underpredicted the maximum daily ozone, particularly during the summer. This is due to several important meteorological parameters, such as cloud fraction and solar radiation, were unavailable for the training period in this study. The CNN model was generally under-trained for forecasting the high ozone peaks during the hot summer days of 2017.

The study also demonstrated that the predictions of the CNN model were generally more accurate (higher IOA and r values with a lower mean bias) for the southern region of the Han River since the topography was more consistent and resulted in a more accurate interpretation of the input parameters. Furthermore, the CNN predictions of daytime ozone concentrations were generally more accurate than those of nighttime concentrations with differences in IOA between 0.05 and 0.30. We attribute this discrepancy to the occurrence of most of the daily maximum input variables during the daytime.

The CNN model not only predicts real-time ozone concentration with favorable statistics but also generates the result within less than a minute of initiating the model. Beyond surveying the advances of utilizing a deep CNN, we illustrated the limitation of such methods for real-time air quality forecasting. For instance, a proper number of input variables (predictors) should be used with a sufficiently large amount of training data. However, if an important predictor of ozone concentration is missing (e.g., cloud fraction and solar radiation), it will influence the sensitivity of the model to the other input variables, which may lead to "misprediction."

The proposed approach in this paper can be applied to and yield a high prediction accuracy for ozone or other pollutants in other metropolitan areas. In addition, the deep learning approach can potentially be used for a multiple-day forecast of air pollution or air quality index. Fast and accurate air quality prediction using the deep CNN model could be used to reduce the adverse health effects of urban air pollution. Given the computational efficiency of the CNN algorithm, deep learning could supplement deterministic models to more rapidly and accurately forecast air pollution concentrations. We expect that this paper will not only provide a more comprehensive understanding of CNNs but also facilitate future research activities and applications within the field of atmospheric sciences.

**Acknowledgments:**
This study was supported by funding from the Department of Earth and Atmospheric Science (EAS Research Grant) of the University of Houston and the National Institute of Environmental Research (NIER).

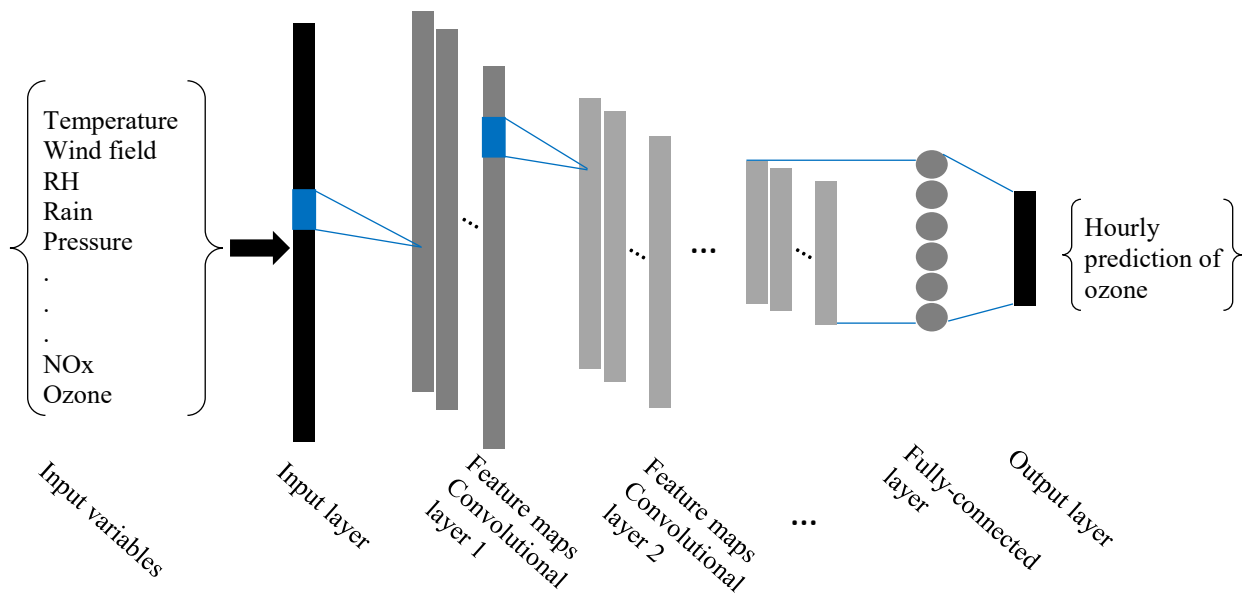

**Fig. 1.** Schematic of the deep CNN in our approach.

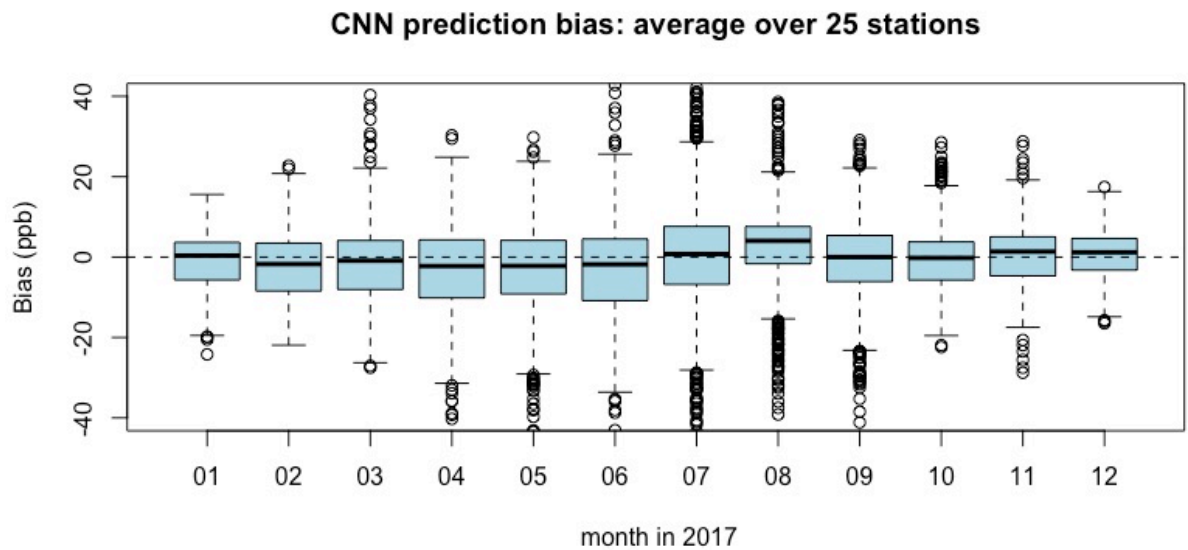

**Fig. 2.** Difference between the ozone observation and the results of CNN ozone prediction systems for all 25 stations in Seoul.

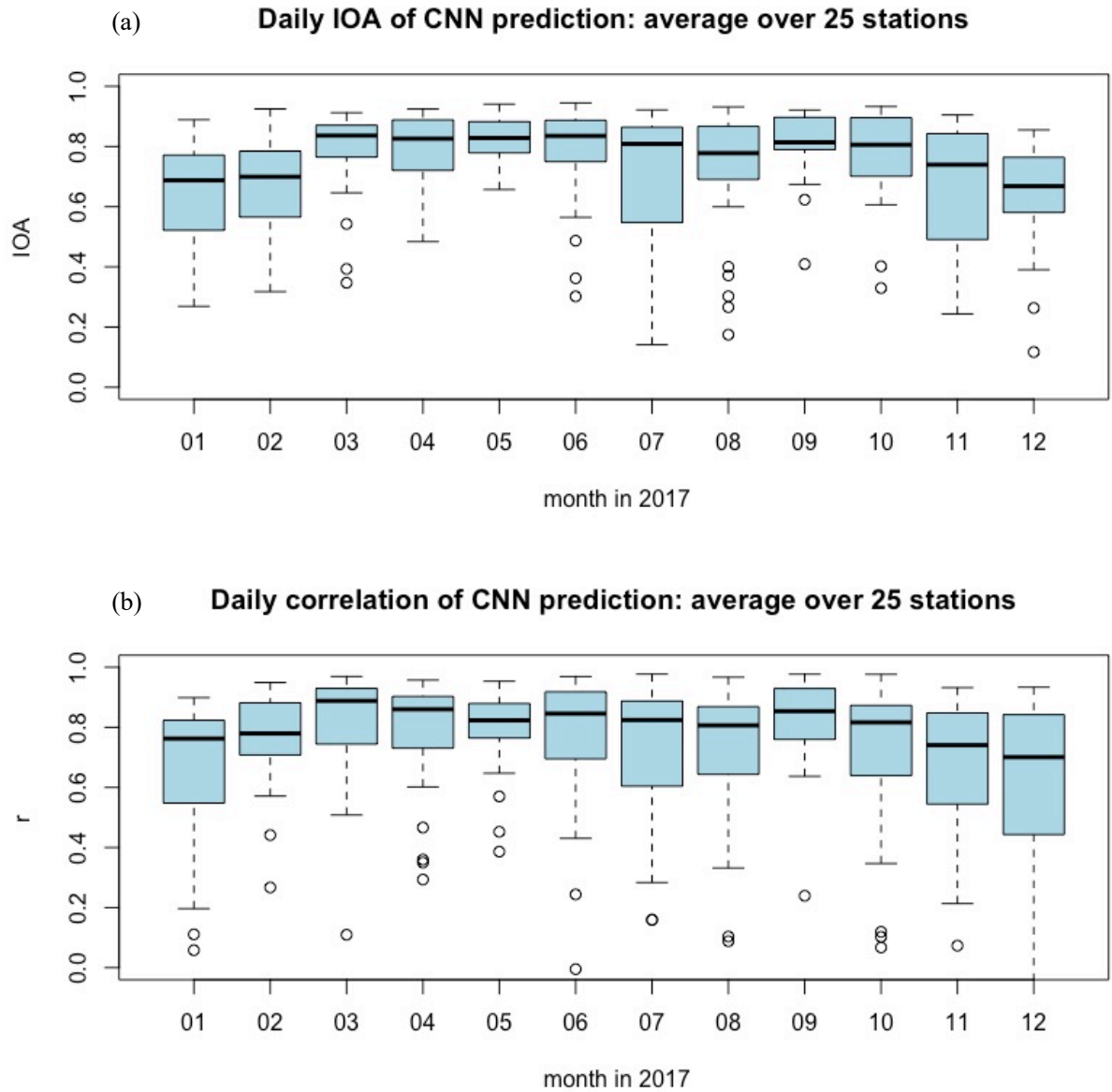

**Fig. 3.** (a) Daily index of agreement and (b) the correlation coefficient of CNN forecasting systems averaged over the 25 stations in Seoul, South Korea.

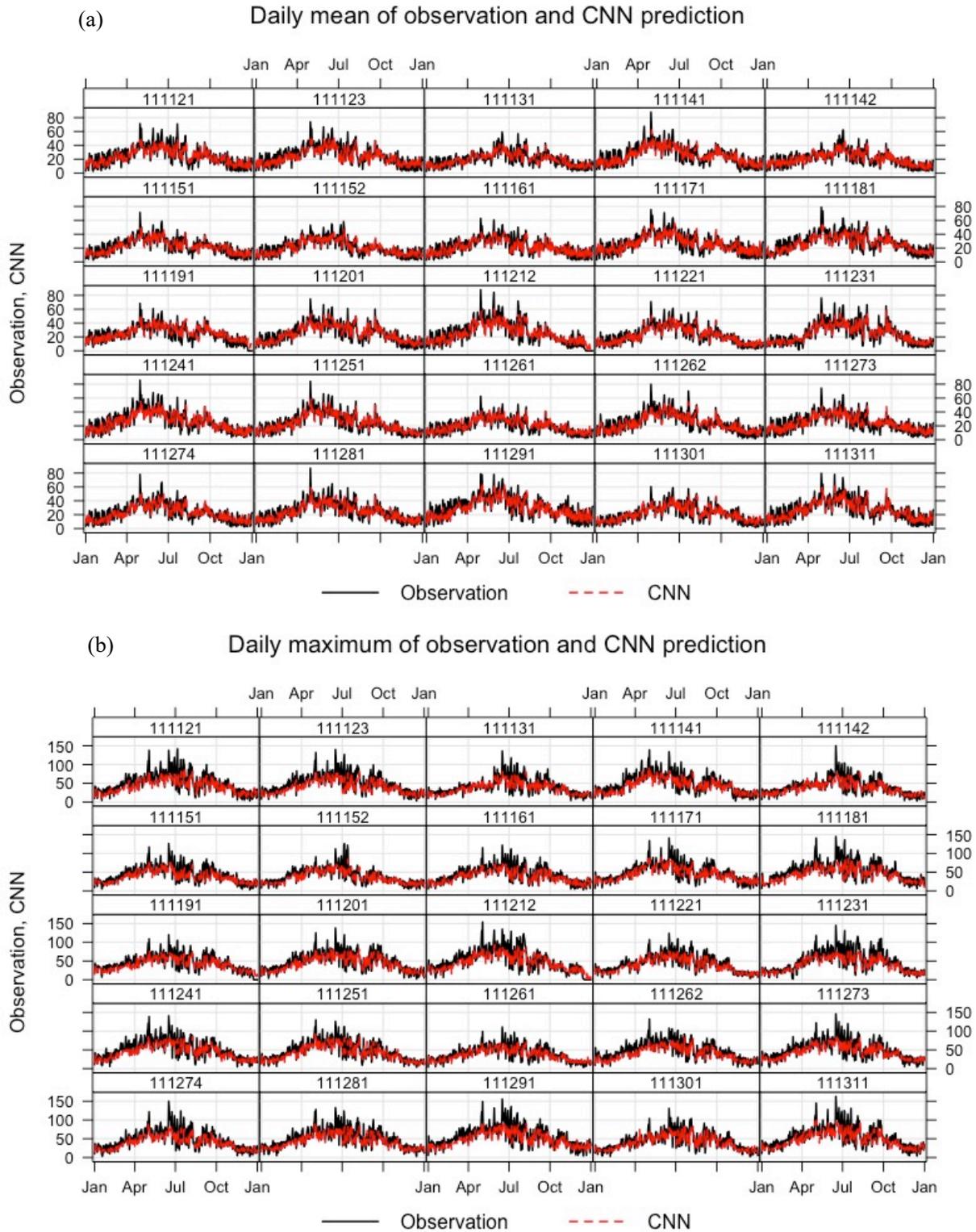

**Fig. 4.** (a) Daily mean and (b) daily maximum of the observations and the CNN prediction for the NIER stations in Seoul.

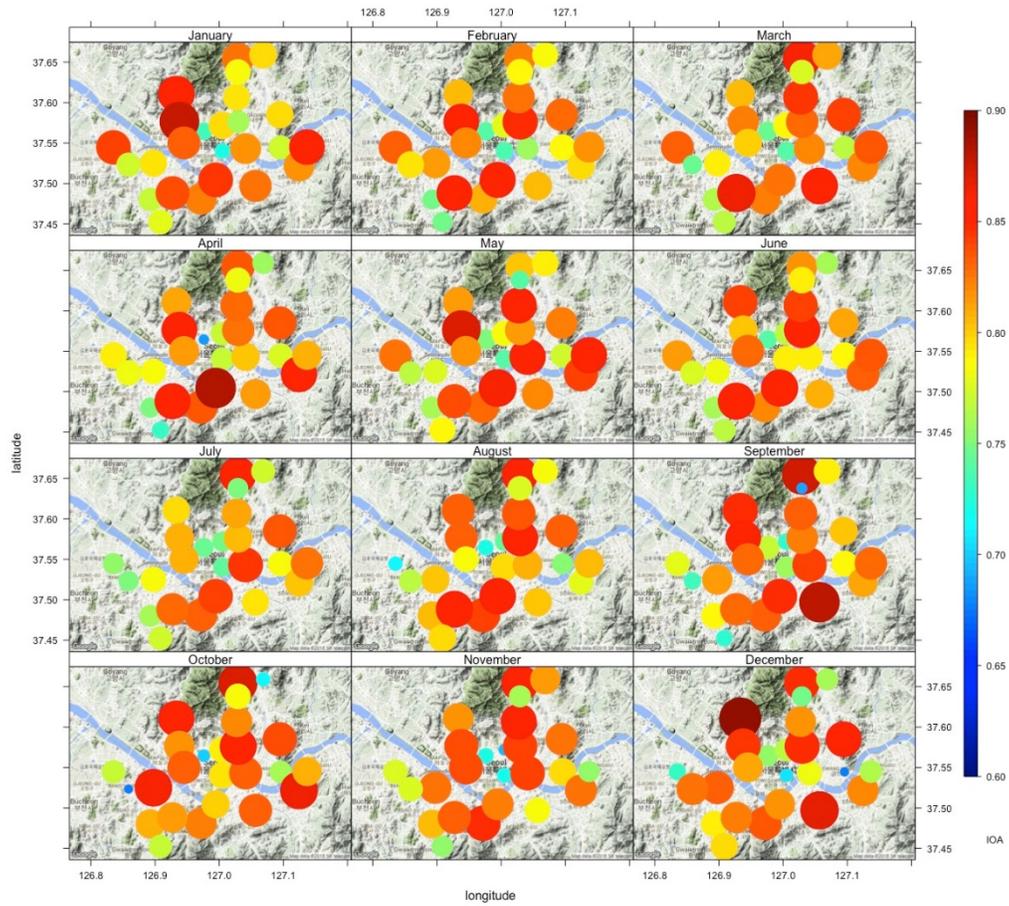

**Fig. 5.** A monthly index of agreement for the NIER stations in Seoul; the larger the size of representing a circle, the higher the value is.

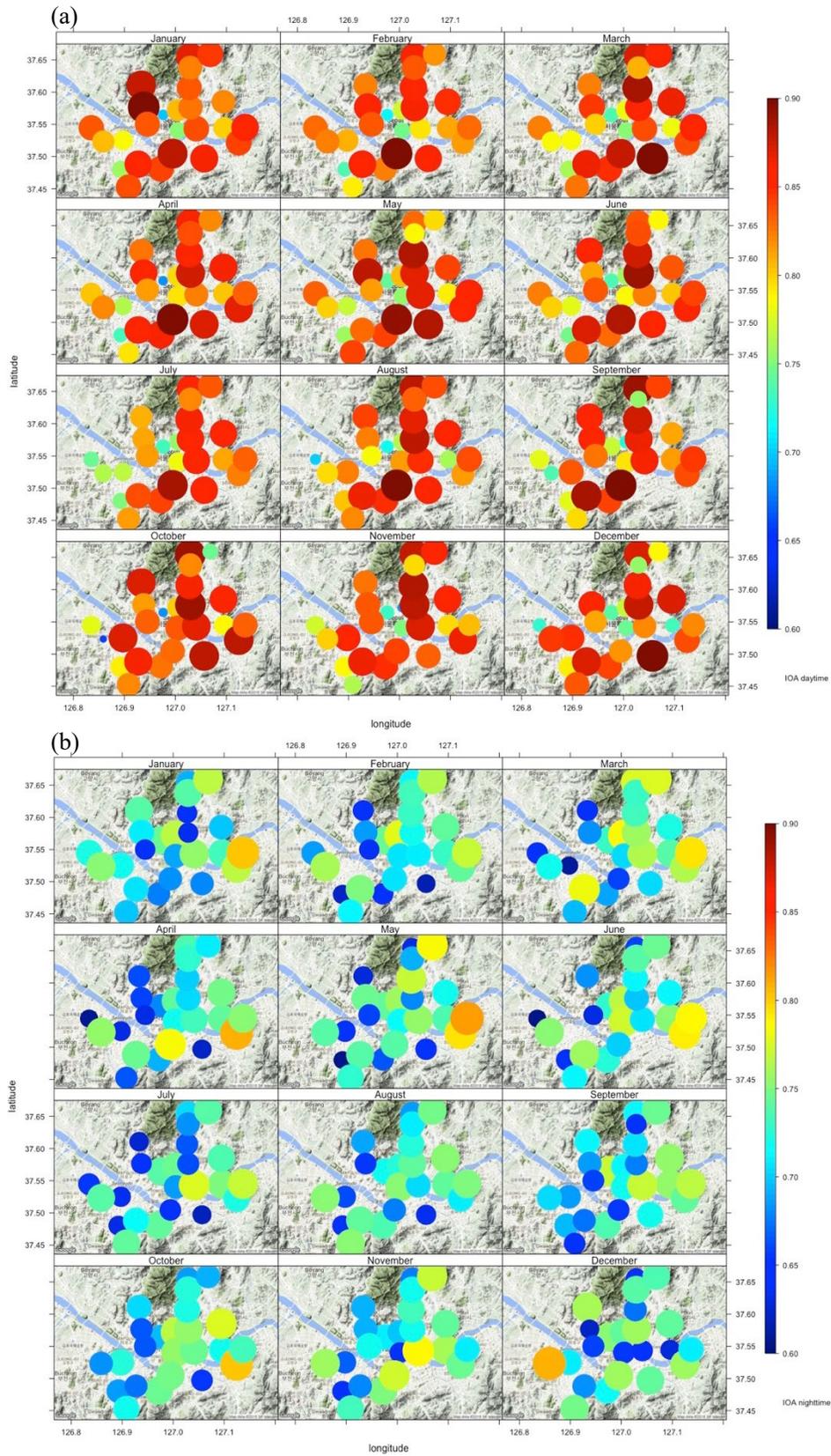

**Fig. 6.** A monthly index of agreement for the NIER stations in Seoul during (a) the daytime, and (b) the nighttime.

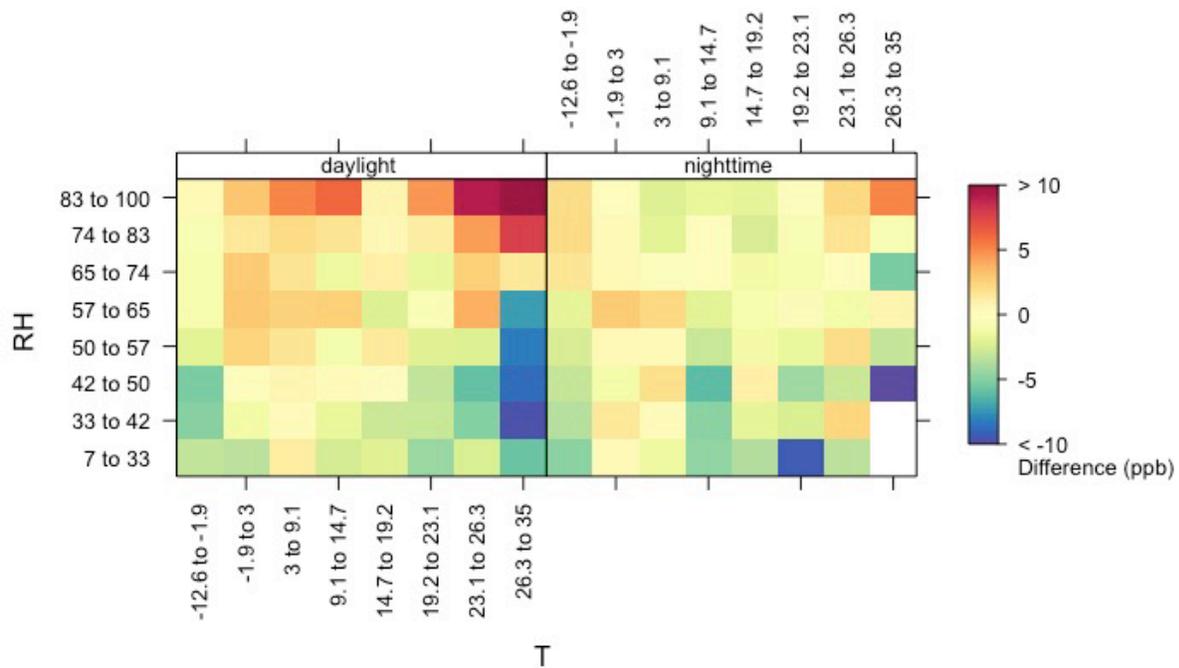

**Fig.7.** Mean bias of the CNN prediction in a categorical comparison between the daytime and the nighttime. The relative humidity is in percentages, and the surface temperature is in degrees Celsius

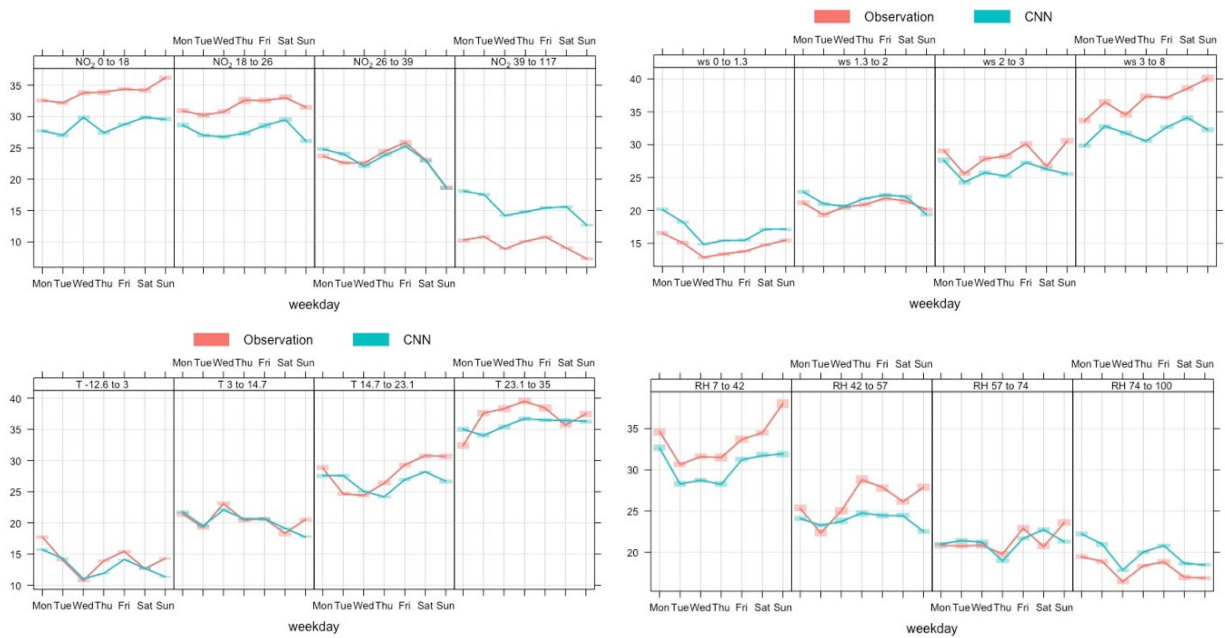

**Fig. 8.** The annual mean of the CNN prediction in a categorical comparison between days of the week. The wind speed is in m/s (upper left), the surface temperature is in degrees Celsius (upper right), $NO_2$ is in ppb (bottom left), and the relative humidity is in percentages (bottom right).

**Supplementary document for "A real-time hourly ozone prediction system using deep convolutional neural network" by Eslami et al.**

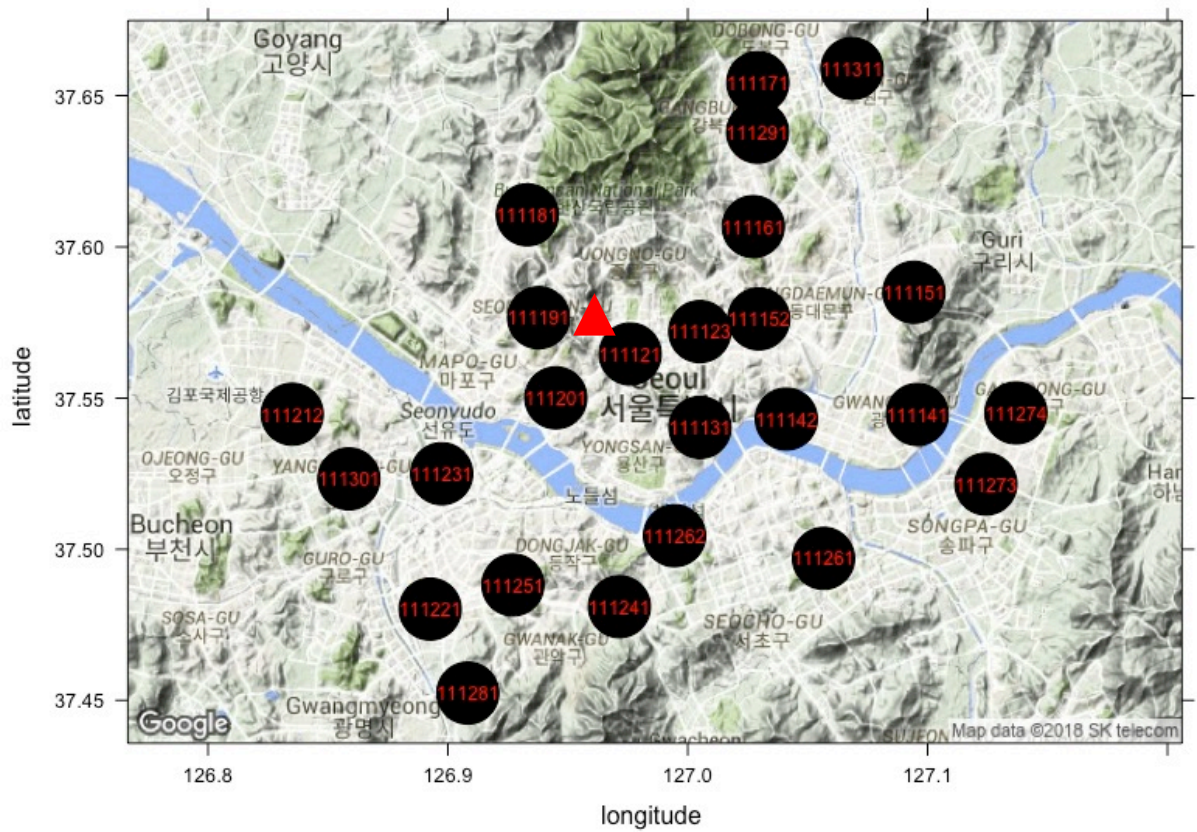

**Fig. S1.** Location of ozone monitoring network by managed by Korea's NIER. The red triangle is the location of meteorology station managed by KMA.

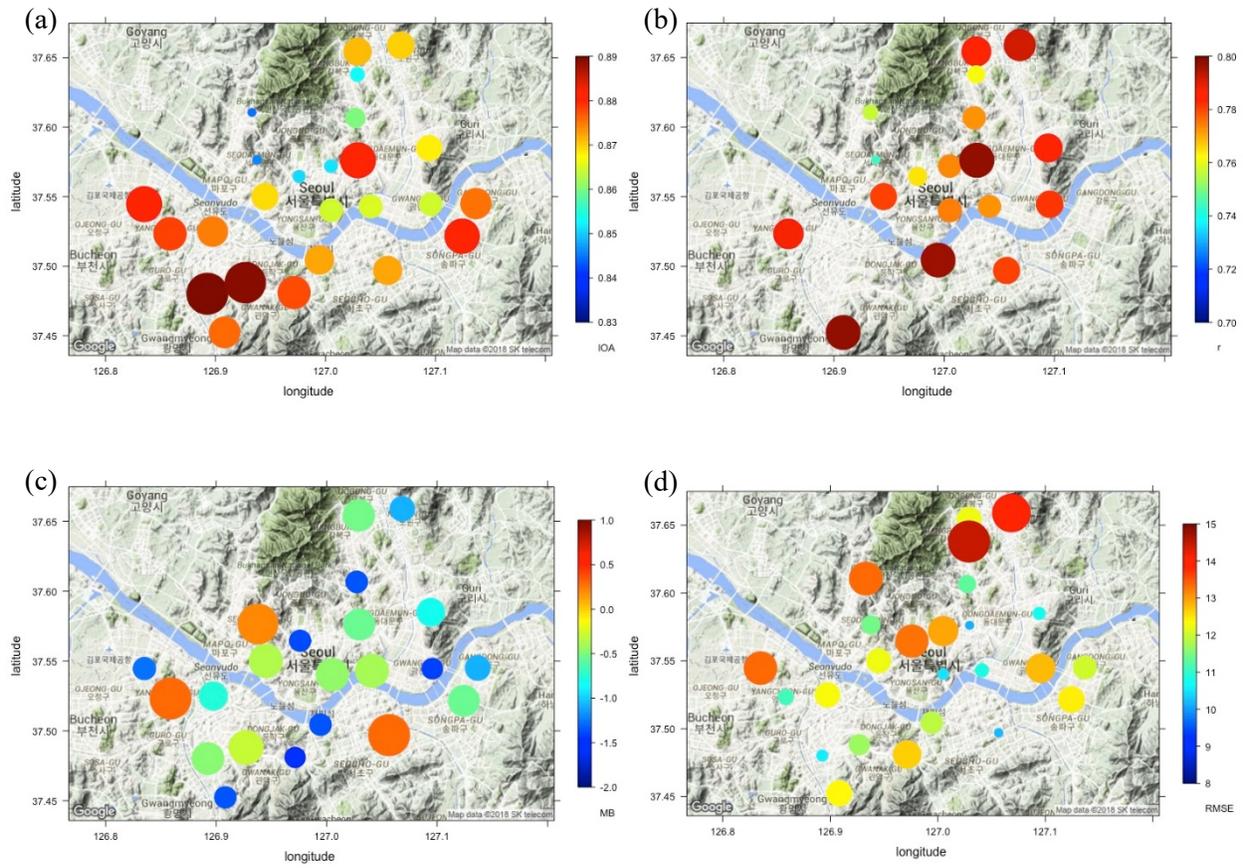

**Fig. S2.** Spatial distribution of (a) Index of Agreement (IOA), (b) Pearson correlation coefficient (r), (c) Mean bias (MB), and (d) root mean squared error (RMSE) of CNN prediction for NIER stations in Seoul.

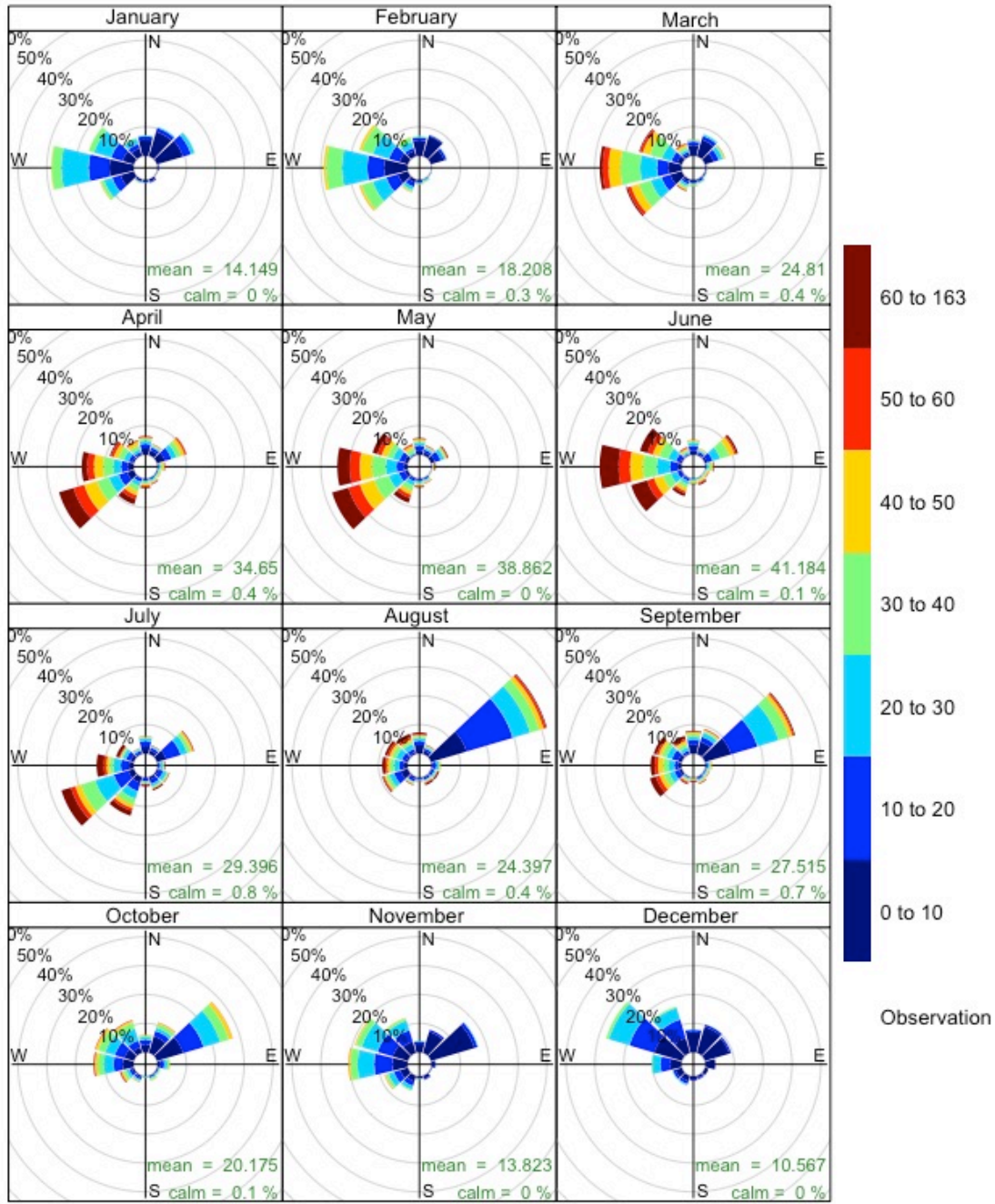

**Fig. S3.** Pollution rose of ozone observation in different month of the year 2017 averaged over all Seoul stations.

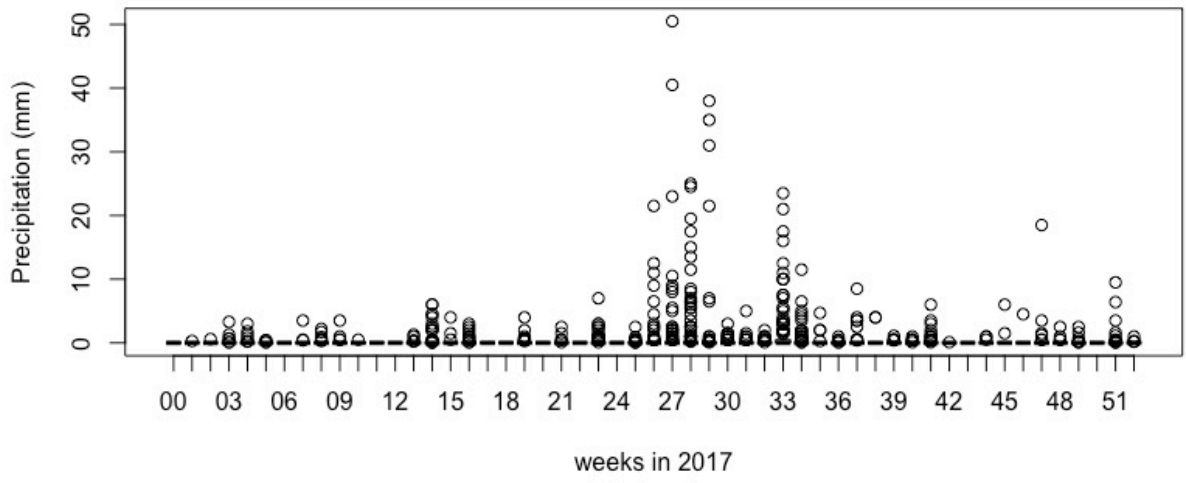

**Fig. S4.** Precipitation levels in different weeks of the year 2017.

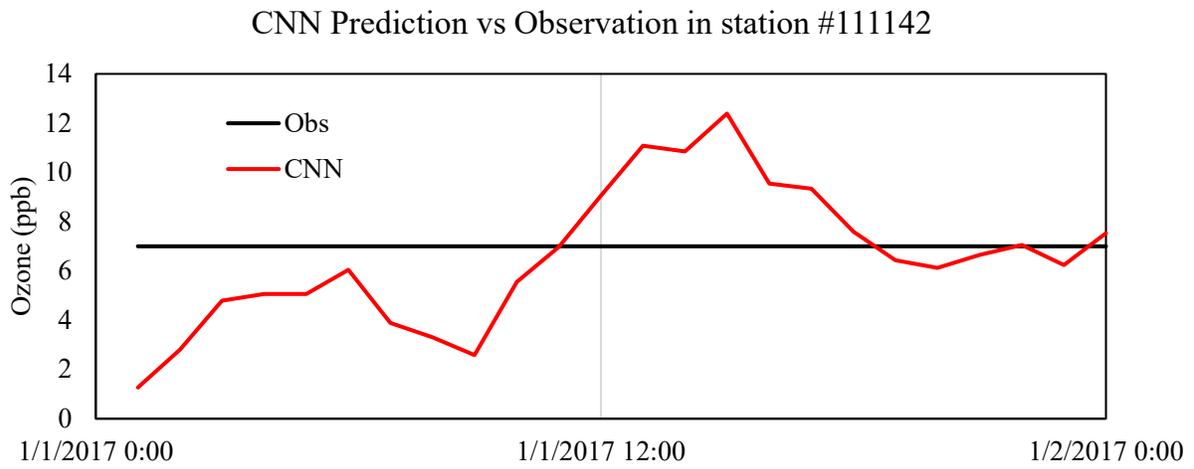

**Fig. S5.** An example of "outlier" prediction by CNN forecasting system.

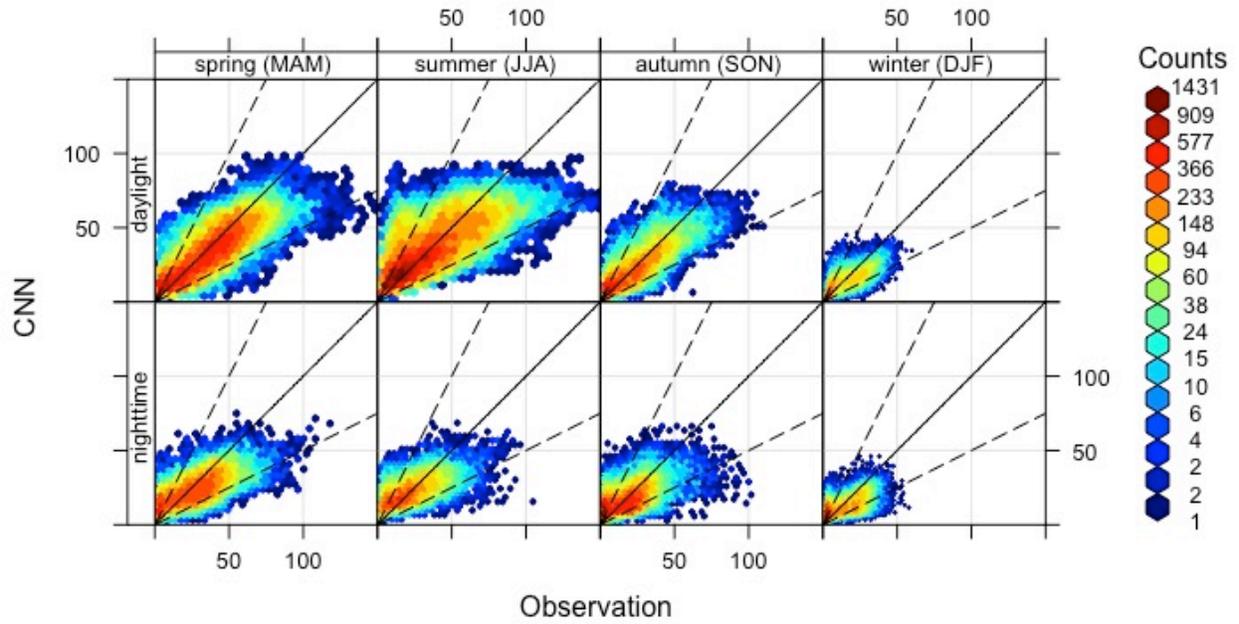

**Fig. S6.** Scatter plot of a comparison between CNN prediction and observation in different seasons averaged over all Seoul stations.

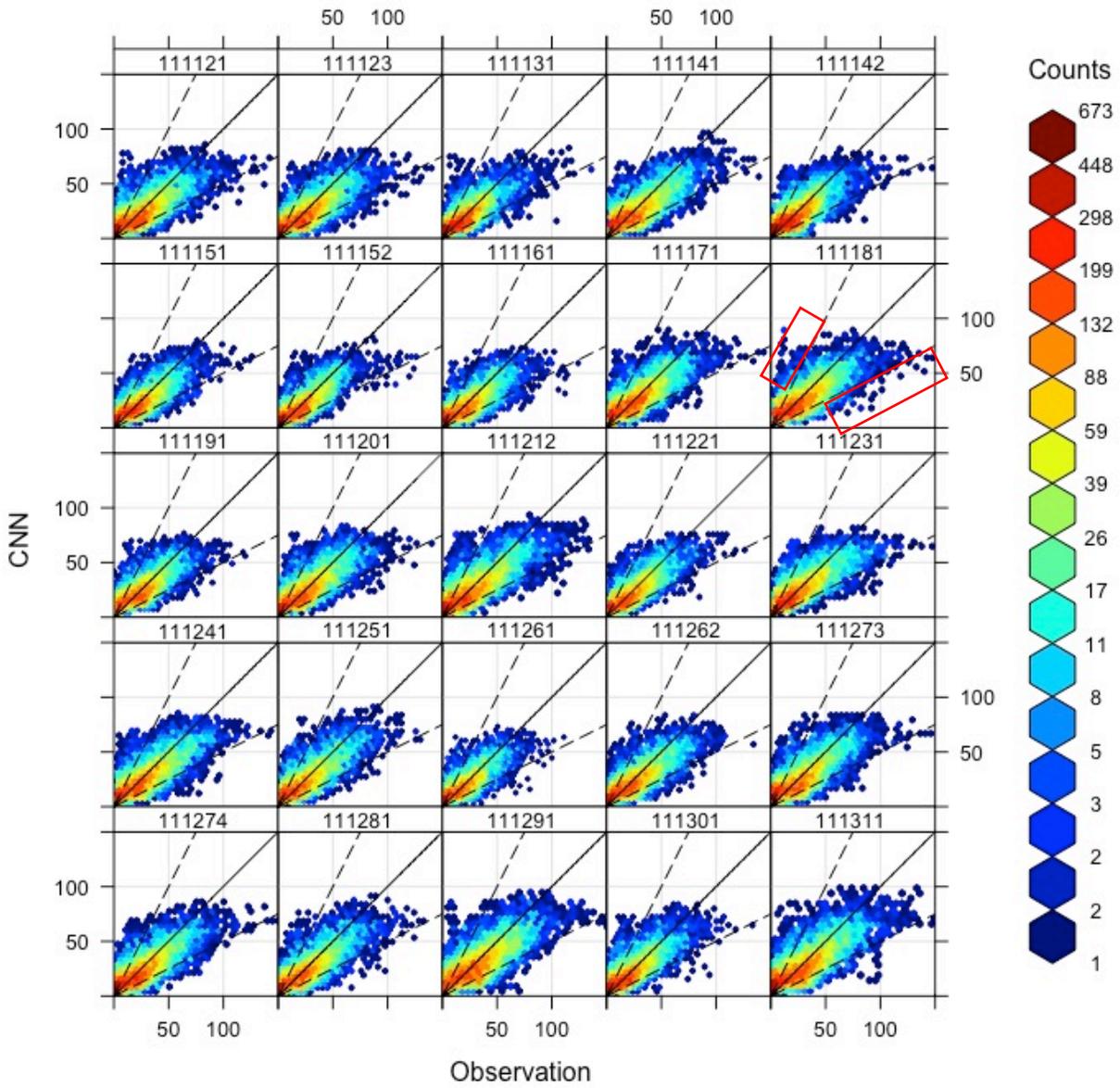

**Fig. S7.** Scatter plot of a comparison between CNN prediction and observation in all Seoul stations for the entire year 2017. An example of misprediction can be seen for Station 111181.

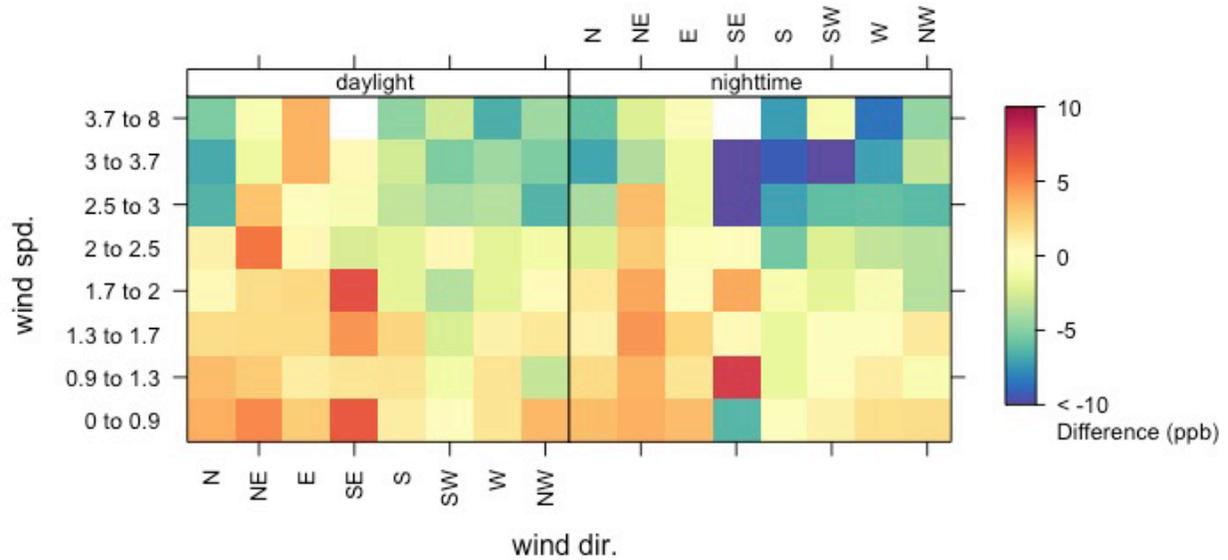

**Fig. S8.** Mean bias of CNN prediction in categorical comparison during daytime and nighttime. The wind speed is in m/s.

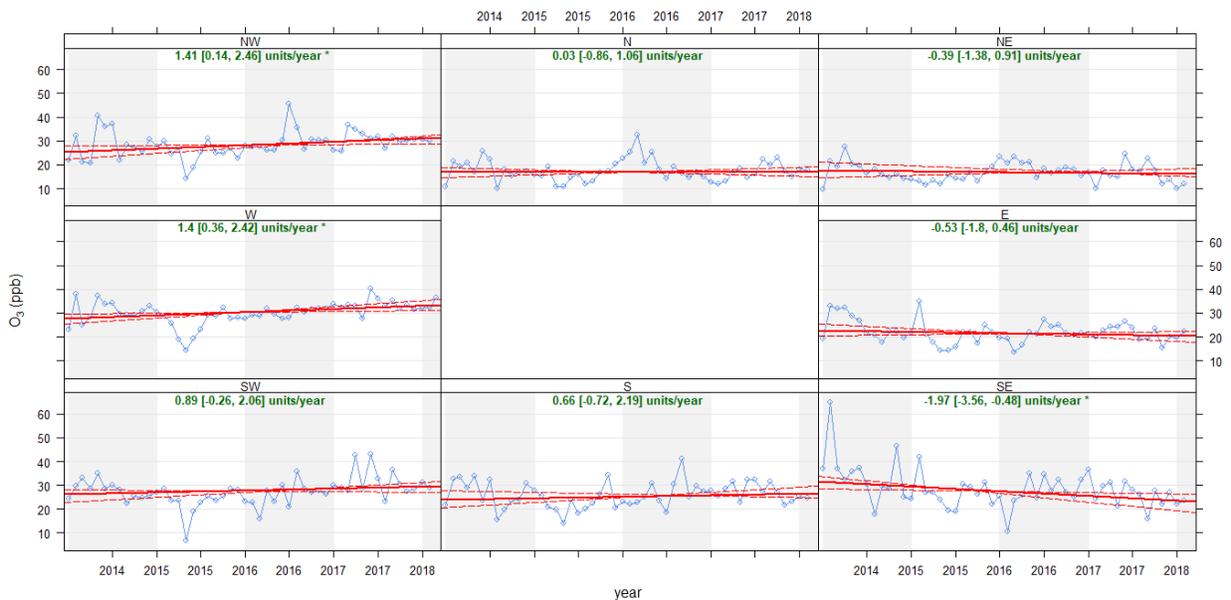

**Fig. S9.** The change in ozone concentration trends in different wind directions in station 111121.

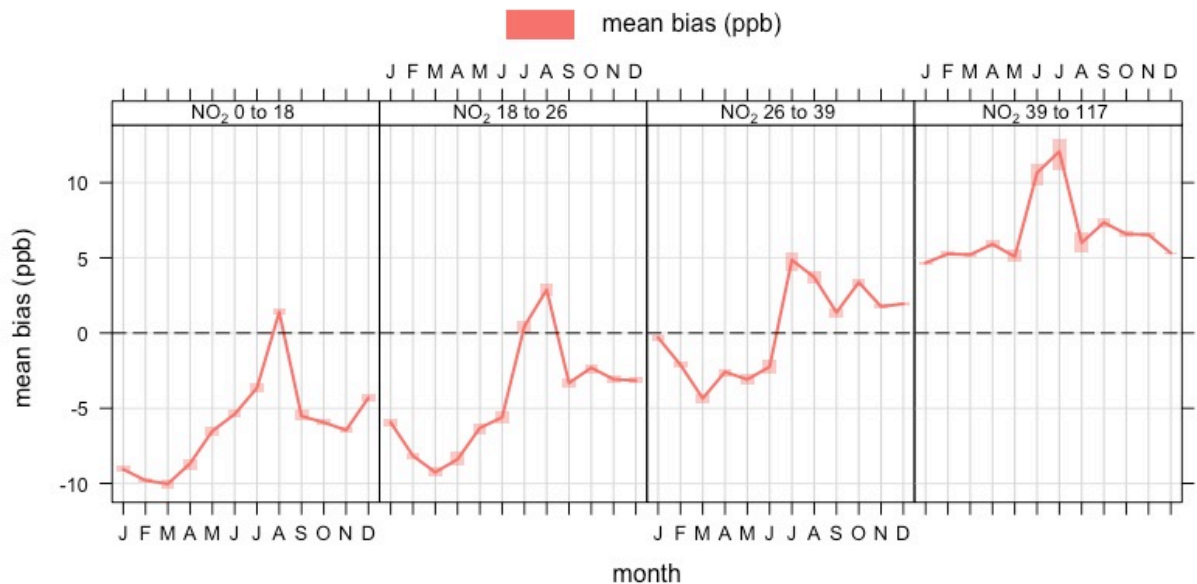

**Fig. S10.** Sensitivity of mean bias of prediction to the different range of $NO_2$ (values in ppb) in different months of the year 2017.

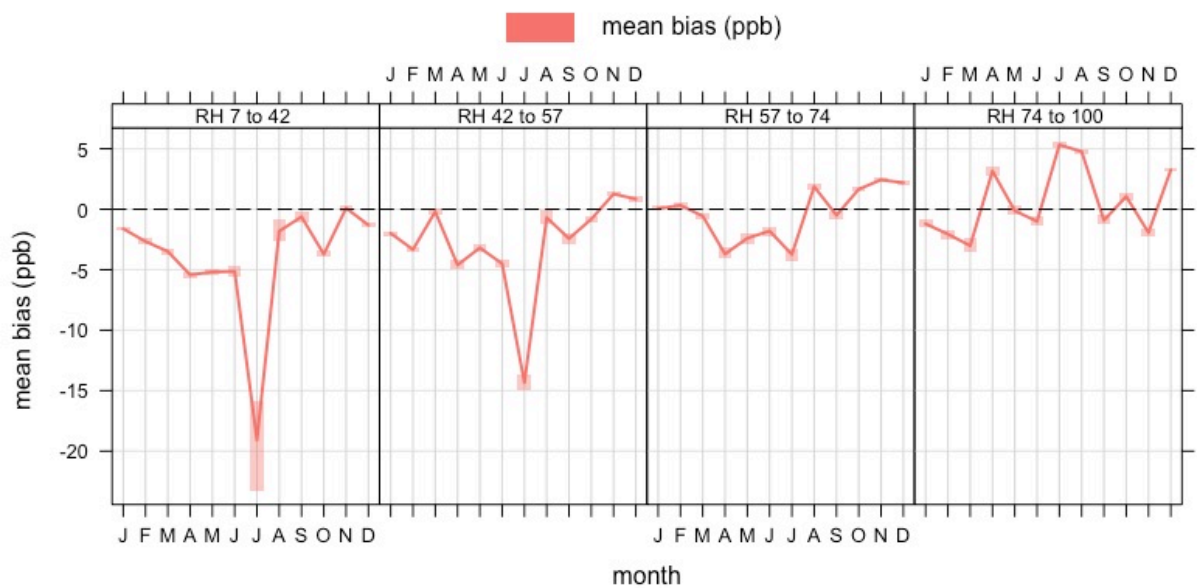

**Fig. S11.** Sensitivity of mean bias of prediction to the different range of relative humidity (RH%) in different months of the year 2017.

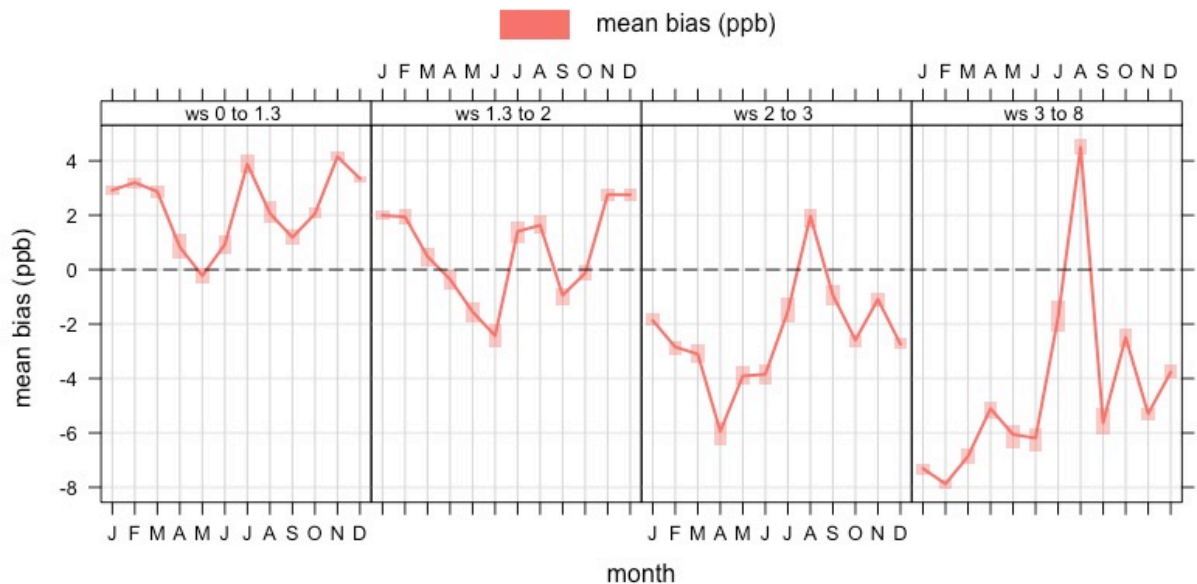

**Fig. S12.** Sensitivity of mean bias of prediction to the different range of wind speed (values in m/s) in different months of the year 2017.